\documentclass[preprint]{jpsj2}

\renewcommand{\theenumi}{( \( \alph{enumi} \) )}
\def\runtitle{The Stochastic State Selection Method 
for Energy Eigenvalues in the Shastry-Sutherland Model }
\def\runauthor{Tomo  \textsc{Munehisa} and Yasuko \textsc{Munehisa}}

\title{%
 The Stochastic State Selection Method \\
for Energy Eigenvalues in the Shastry-Sutherland Model }

\author{%
Tomo  \textsc{Munehisa} and
Yasuko \textsc{Munehisa}
}

\inst{%
 Faculty of Engineering, Univ. of Yamanashi, Kofu 400-8511
}
\recdate{\today}

\abst{%

We apply a recently developed stochastic method to the
Shastry-Sutherland model on $4 \times 4$ and $8 \times 8$ lattices.
This method, which we call the Stochastic State Selection Method here,
enables us to evaluate expectation values of powers of the Hamiltonian 
with very limited number of states.
In this paper we first apply it to the $4 \times 4$ Shastry-Sutherland
system, where one can easily obtain the exact ground state, in order to 
demonstrate that the method works well for this frustrated system. 
We numerically show that errors of the evaluations depend much on 
representations of the states and that the restructured representation 
is better than the normal one for this model.
Then we study the $8 \times 8$ system to estimate energy eigenvalues of 
the lowest $S=1$ state as well as of the lowest excited $S=0$ state, 
where $S$ denotes the total spin of the system. The results, which 
are in good accordance with our previous data obtained by the Operator
Variational method, support that an intermediate spin-gap phase exists 
between the singlet dimer phase and the magnetically ordered phase. 
Estimates of the critical coupling and of the spin gap 
for the transition from the dimer phase to the intermediate phase
are also presented.}

\kword{%

quantum spin, large size, numerical calculation, Monte Carlo,
Shastry-Sutherland
}

\begin{document}
\sloppy
\maketitle
      
\section{Introduction}

Considerable effort has been exerted to develop effective methods to 
calculate various quantities in quantum spin 
systems\cite{dagetal,book1,book2}. 
A new method we have proposed in a previous paper\cite{mm} is a kind 
of Monte Carlo approach, where stochastic variables play an important
role. It is completely different from the conventional quantum Monte 
Carlo methods which employ random walks or importance samplings\cite{book2}.
In our method, which we call the Stochastic State Selection (SSS) 
method hereafter, random variables are used to reduce the number of states
in the vector space which is huge for most systems of large sizes.
To generate these random variables we introduce a probability function 
named {\em on-off} probability function, which ``{\em switches off}'' 
a number of states in the vector space so that 
we can calculate approximate expectation values of powers of 
the Hamiltonian from a small number of the ``{\em on}'' states.
Repeating this process with a new set of random variables 
we can obtain averaged expectation values which are very close to the 
exact ones. 

The fundamental idea in the SSS method is deeply connected with the 
variational approach to quantum spin systems, where one searches 
for an effective basis, a set of states which can be treated 
within the limit of computer facilities.
In the variational approach there are two essential steps: 
increasing the number of states in the wave function 
under current investigation, and then reducing it. 
In the former step one usually operates the Hamiltonian to the wave
function, which is a mathematically established procedure.
How to formulate the latter step, on the other hand, is quite controversial.
In early works\cite{truncs} states with small
coefficients are dropped, but the resultant wave functions are not 
satisfactory.
Sampling the states using the Monte Carlo techniques has also been  
suggested in recent works\cite{imadas, sorl}.
In the SSS method one can reduce the number of states based on
a {\em mathematically justified} way. It is this remarkable property that
makes us to believe the method is worth investigating.

In the previous paper we described the SSS method and applied it to the
two-dimensional spin one-half Heisenberg model.  
In this paper we demonstrate that it can be
applied to the system suffering from the negative sign problem, too.
A concrete example is the Shastry-Sutherland (SS) 
model\cite{ss,miya,fuku,albr,koga,knett,chung,zheng,lauchli,ov} 
on a $4 \times 4$ lattice, whose results reproduce the exact values in good 
precision. Next we try to obtain physical quantities, 
the energy eigenvalues with fixed values of total spin $S$, 
for the $8 \times 8$ lattice changing the coupling ratio
across the critical region.

In the next section we briefly explain the SSS method.  
After defining the on-off probability function and the random choice matrix, 
we describe how we calculate expectation values of $\hat H ^L$, 
$\hat H$ being the Hamiltonian of the system under study.
Section 3 is a short description of the SS model, where
we explain the Hamiltonian and several properties of this model. 
In section 4 numerical results on a $4 \times 4$ lattice near the
critical point are presented for two representations, 
{\em normal} and {\em restructured}\cite{mmptp}.
Deviations of the data here indicate that 
the restructured representation is more effective 
for the coupling ratio we study in this section, which supports
the picture that the orthogonal dimers dominate here.
Based on these discussions we try in section 5 to obtain physical 
quantities on a $8 \times 8$ lattice, 
the energy eigenvalues with the total spin $S=0$ and $S=1$, 
in the intermediate region of the coupling ratio. 
As noted previously\cite{mm},
there are two model-dependent problems to be solved.
One of them is how we generate a {\em good} approximate state 
$\mid \psi_{\rm A} \rangle$.
Another is how we extract eigenvalues from the expectation values 
$\langle \psi_{\rm A} \mid \hat H ^L \mid  \psi_{\rm A} \rangle$.
In addition to the calculations of expectation values for
the restructured representation, extensive discussions on 
these subjects are given in this section. 
We will show that we successfully obtain the results on the energy 
eigenvalues which indicate a phase transition 
from the singlet dimer phase to an intermediate spin-gap phase.   
Finally, the last section is devoted to summary and discussions.
 
\section{Stochastic State Selection Method}

Here we give a brief description of the SSS method\cite{mm}. 
We define the on-off probability function by
\begin{eqnarray}
 P(\eta) \equiv \frac{1}{a}\delta(\eta -a) +(1- \frac{1}{a})\delta(\eta),
\label{on-off}
\end{eqnarray}
where $a$ is a constant which is greater than or equal to 1.
It is clear by definition that 
\begin{eqnarray}
   \langle \! \langle  1 \rangle \! \rangle  & \equiv & \int_0^\infty P(\eta) d \eta =  1 , \\  
\langle \! \langle   \eta \rangle \! \rangle &\equiv & \int_0^\infty \eta  P(\eta) d \eta = 1,
\label{eta1}
 \\
  \langle \! \langle   \eta^2 \rangle \! \rangle  &\equiv & \int_0^\infty \eta^2  P(\eta) d \eta = a ,
\label{eta2}
\end{eqnarray}
where $\langle \! \langle  \rangle \! \rangle$ denotes the
statistical average.

Let us expand a state $\mid  \psi \rangle$ by a basis \{ $\mid  i \rangle$\}
\begin{eqnarray}
\mid  \psi \rangle = \sum_{i=1}^{N_V} \mid  i \rangle c_i,
\label{expnd0} 
\end{eqnarray} 
where $N_V$ denotes the size of the full vector space, 
and define the expectation values $ E(L)$ ($L = 1,2,\cdots$) as  
\begin{eqnarray}
 E(L) \equiv \langle \psi \mid  \hat H^L \mid  \psi \rangle \ .
\end{eqnarray}
We then introduce the random choice matrix, 
an $N_V \times N_V$ diagonal matrix,
\begin{eqnarray}
M_{ \{ \eta \}} \equiv 
\left( \begin{array}{cccc} \eta_1 & 0 & \cdots & 0 \\
0 & \eta_2 & \cdots & 0  \\
\cdots & \cdots & \cdots & \cdots \\
 0 & 0 & \cdots & \eta_{N_V} \end{array}  
\right) \ .
\label{rdmcmx}
\end{eqnarray} 
Each random variable $\eta_i$ in (\ref{rdmcmx}) is determined 
by the on-off probability function
\begin{eqnarray}
  P_i(\eta_i) & = &
\frac{1}{a_i}\delta(\eta_i -a_i) +(1- \frac{1}{a_i})\delta(\eta_i) ,
\label{pi}
\end{eqnarray}
where 
\begin{eqnarray}
a_i \equiv \left\{ \begin{array}{ll}
{\rm max}(1,\epsilon/ |c_i| ) & ({\rm if} \ c_i \neq 0) \\
\epsilon/\delta       & ({\rm if} \ c_i = 0) \ ,
\end{array} \right.
\label{ai} 
\end{eqnarray}
with given constants $\epsilon $ and $\delta$ (0 $<$ $\delta$ $<$ $\epsilon $).
Using $(L+1)$ independent random choice matrices 
$M_{\{ \eta ^{(m)}\}} = diag.\{ \eta_1^{(m)}, \eta_2^{(m)},\cdots, 
\eta_{N_V}^{(m)}\}$\cite{foot0} we define  
\begin{eqnarray}
  E_{\{\eta\}}(L) \equiv \langle \psi \mid M_{\{ \eta^{(L+1)}\}} \hat H 
M_{\{ \eta ^{(L)}\}} \hat H M_{\{ \eta ^{(L-1)}\}} \cdots \hat H 
M_{\{ \eta ^{(1)}\}} \mid  \psi \rangle .
\label{eldef}
\end{eqnarray}
It is easy to see that
$\langle \! \langle E_{\{\eta \}}(L) \rangle \! \rangle =  E(L)$.
Note that by operating $M_{\{\eta^{(m)}\}}$ 
we can obtain a truncated vector space for 
$\hat H M_{\{ \eta ^{(m-1)}\}} \cdots \hat H 
M_{\{ \eta ^{(1)}\}} \mid  \psi \rangle$,
since $\eta_i^{(m)}$ is {\em zero} with the probability $1-1/a_i$.  
Also note that the variances of $\eta_i^{(m)}$ are the same for all $m$ 
because both $\langle \! \langle \eta_i^{(m)}\rangle \! \rangle $
and $\langle \! \langle [ \eta_i^{(m)} ]^2 \rangle \! \rangle $ do not
depend on $m$.

In numerical studies we measure
\begin{eqnarray}
 \ \langle \! \langle E_{{\{\eta \}}}(L) \rangle \! \rangle _{{\rm smpl}} &\equiv&
\frac{1}{n_{{\rm smpl}}} \sum_{k=1}^{n_{{\rm smpl}}} E_{{\{\eta \}_k}}(L) ,
\label{wem4}
\\
 \rho_{{\{\eta \}}}^2(L) & \equiv &
 \ \langle \! \langle [E_{{\{\eta \}}}(L) ]^2 \rangle \! \rangle _{{\rm smpl}}
 - \ \langle \! \langle E_{{\{\eta \}}}(L) \rangle \! \rangle _{{\rm smpl}} ^2 \ ,
\label{wem5}
\end{eqnarray}
using $n_{{\rm smpl}}$ samples. 
The error of $E_{{\{\eta \}}}(L) $ is evaluated by
\begin{eqnarray}
Er(L) \equiv 2\sqrt{ \frac {\ \rho_{{\{\eta \}}}^2(L)}{n_{{\rm smpl}}}}.
\label{wem6}
\end{eqnarray}
The fact that the number of non-zero
components much reduces by operating $M_{\{ \eta ^{(m)}\}}$ to  
$\hat H M_{\{ \eta ^{(m-1)}\}} \cdots \hat H 
M_{\{ \eta ^{(1)}\}} \mid  \psi \rangle $ is essential to numerical
calculations. Let us denote by
$\langle \! \langle  N_b(m) \rangle \! \rangle _{{\rm smpl}}$ and
$\langle \! \langle  N_a(m) \rangle \! \rangle _{{\rm smpl}}$
the number of non-zero coefficients we count
in the measurement $before$ and $after$ we operate $M_{\{ \eta ^{(m)}\}}$ to
$\hat H M_{\{ \eta ^{(m-1)}\}} \cdots \hat H 
M_{\{ \eta ^{(1)}\}} \mid~\psi \rangle $, respectively. 

\section{Shastry-Sutherland Model}
As is discussed in several papers
\cite{miya,fuku,albr,koga,knett,chung,zheng,lauchli,ov},
one of recent fascinating topics in two-dimensional quantum spin systems is
the Shastry-Sutherland (SS) model\cite{ss},
the model of orthogonal dimers with the intra-dimer coupling $J \ ( > 0)$
and the inter-dimer coupling $J'\ ( > 0)$.

In this model a spin is located at the sites
$$ (2n_x a + \frac{d}{2} , 2n_y a+ \frac{d}{2} ), \ \ \
 (2n_x a - \frac{d}{2} , 2n_y a- \frac{d}{2} ), $$
$$ ((2n_x+1) a + \frac{d}{2} , (2n_y+1) a- \frac{d}{2} ), \ \ \
 ((2n_x+1) a - \frac{d}{2} , (2n_y+1) a+ \frac{d}{2} ),$$
on the $2N \times 2N$ lattice in Fig. 1, 
where $n_x =0,1,\cdots ,N-1$ and $n_y =0,1, \cdots ,N-1$,
$2a$ is the unit distance between dimers
and $\sqrt{2}d$ is the distance of spins in a dimer. 
The Hamiltonian is given by
\begin{eqnarray*}
\hat{H} & = & \frac{1}{4} J \sum_{n_x,n_y=0}^{N-1}  
\{  h_a(n_x,n_y)+h_b(n_x,n_y) \}  \\
& + & \frac{1}{4} J' \sum_{n_x,n_y=0}^{N-1}
 \{ h_1(n_x,n_y)+h_2(n_x,n_y)+h_3(n_x,n_y)+h_4(n_x,n_y) \} ,
\end{eqnarray*}
where the partial Hamiltonians are, as given in ref.17,  
\begin{eqnarray*}
h_a(n_x,n_y) & \equiv & \mbox{\boldmath $\sigma$}
(2n_x a + \frac{d}{2} , 2n_y a+ \frac{d}{2} )\cdot  
\mbox{\boldmath $\sigma$}(2n_x a - \frac{d}{2} , 2n_y a- \frac{d}{2} ),
\end{eqnarray*}
and so on with \mbox{\boldmath $\sigma$}$(x,y)$ denoting the Pauli matrix 
at the location $(x,y)$. 

This model provides a nontrivial system which relates an
exactly solvable spin-gap model obtained in the $J'\rightarrow 0$ limit
to a square lattice model where intra-dimer interaction vanishes ($J = 0$).
It is known that the exact ground state of the SS model is the direct
product of the singlet dimers when $J'\le J'_c \sim 0.68J$\cite{koga}, while
for sufficiently large $J'/ J$ the system is in the
magnetic order phase with the N\'{e}el ground state.
It is almost doubtless that another phase exists between these two 
phases, but the location and the properties of this phase are not 
definite yet\cite{albr,koga,knett,chung,zheng,lauchli,ov}. 
This phase is suggested to be gapless or nearly so in
refs.13 and 15, while authers in refs.11, 12, 14, 16 and 17 conclude 
it is gapped. Numerical data for the spin-gap is presented only in
refs.12 and 17. 
 
In study of these phases a fundamental quantity to be examined is
the difference between the energy of the singlet-dimer state
$E_0$ and the lowest energy $E$ as a function of the coupling $J'/J$, 
when the momentum \mbox{\boldmath $p$}=$(0,0)$,
the total spin $S=j \ (j= 0, 1)$ and the $I_+$ parity is $even$ or $odd$, 
where the $I_+$ parity is the parity for the $x$-$y$ reflection 
$I_+ \mbox{\boldmath $\sigma$} (x,y)I_+^\dagger = $ 
\mbox{\boldmath $\sigma$}$(y,x)$.

\section{Numerical results on a 4$\times$4 lattice}

In this section we show the numerical results obtained  
on a small lattice in order to demonstrate that we can apply the SSS 
method to the SS model with $J' \sim J'_c$, where the negative sign
problem is serious, without any difficulty.

It is quite easy to calculate the exact ground state 
$\mid  \psi _{\rm E} \rangle$ of the model on the 4$\times$4 lattice 
using the Lanczos techniques. 
We therefore do not need to pursue any trial state here. We just measure 
\begin{eqnarray}
E_{{\rm E}\{\eta\}}(L) \equiv \langle \psi_{\rm E} \mid M_{\{ \eta^{(L+1)}\}} \hat H 
M_{\{ \eta ^{(L)}\}} \hat H M_{\{ \eta ^{(L-1)}\}} \cdots \hat H 
M_{\{ \eta ^{(1)}\}} \mid  \psi_{\rm E} \rangle
\end{eqnarray}
($L=1,2,\cdots$) and compare them with $E_{{\rm E}}(L)= E^L$, where $E$
denotes the eigenvalue of $\hat H $, 
namely $\hat H \mid \psi_{\rm E} \rangle = \mid \psi_{\rm E} \rangle E$.

In this section we fix the coupling ratio $J'/J = 0.7$, a value 
slightly above the critical point, for which $E/J=-6.3502$. 
Here we calculate 
$\langle \! \langle E_{{\rm E}{\{\eta \}}}(L) \rangle \! \rangle _{{\rm smpl}}$ 
using two representations. One of them is the normal one, where the
state on each site of the lattice is represented by the $z$ component of
the spin sitting on the site. In another representation we named 
$restructured$, two spins connected with the coupling $J$ are rearranged
in three triplets and one singlet states\cite{mmptp}.
The total number of non-zero coefficients in expansion of 
$\mid  \psi_{\rm E} \rangle $ is 12,870 (8,565) in the normal (restructured) 
representation. 

Table I shows 
our results up to $L=10$ obtained from $10^4$ samples with 
$\epsilon = 0.1$, together with the exact values. 
Note that $\delta$ in (\ref{ai}) is not necessary for 
$\mid  \psi _{\rm E} \rangle$. 
We see that the results using the restructured representation are in
good agreement with the exact ones. Especially for small values of $L$
the agreement is excellent. The relative errors are in the range from 
$0.11\%$ ($L=1$) to $2.8 \%$ ($L=10$). The results from the normal
representation are also compatible with the exact values. We, however, 
observe that the statistical errors are much larger than those in the
restructured representation except for very low $L$.  
The relative error is $0.12\%$ for $L=1$ but it rapidly increases as $L$
grows to reach $113\%$ $(L=9)$, where the evaluated value is not meaningful. 
For larger values of $\epsilon$ such tendency is more outstanding. 
With $\epsilon = 0.2$ and $n_{{\rm smpl}} = 10^4$, for example, 
we cannot obtain reliable values of
$\langle \! \langle E_{{\rm E}{\{\eta \}}}(L) \rangle \! \rangle _{{\rm smpl}}$ 
$(L \geq 6)$ in the normal representation, while the results from the 
restructured representation are satisfying up to $L=10$. 
When the number of samples is increased we are in principle to obtain 
satisfactory evaluations even in the normal representation. In fact, we
observe that for $10^5$ samples with $\epsilon = 0.1$ the statistical
errors become smaller than those for $10^4$ samples except for $L=10$.  
$\langle \! \langle E_{{\rm E}{\{\eta \}}}(9) \rangle \! \rangle _{{\rm smpl}} / J^9$
and $\langle \! \langle E_{{\rm E}{\{\eta \}}}(10) \rangle \! \rangle _{{\rm smpl}}
/ J^{10}$ for $10^5$ samples are $-(0.189 \pm 0.066)\times 10^8$ and 
$-(0.120 \pm 0.088)\times 10^9$, respectively.
  
Table II presents the numbers of non-zero coefficients measured in the
same calculations with $10^4$ samples, 
which is to show how the number of non-zero 
coefficients decreases by each operation of the random choice matrix.  
In both representations we see that 
$  \langle \! \langle  N^{\rm E}_a(L) \rangle \! \rangle _{{\rm smpl}}$, which
denotes the number of the ``{\em on}'' states after the operation, is
less than a tenth part of 
$  \langle \! \langle  N^{\rm E}_b(L) \rangle \! \rangle _{{\rm smpl}}$, the number
of non-zero coefficient before the operation. In addition, the data
suggest that both $  \langle \! \langle  N^{\rm E}_a(L) \rangle \! \rangle _{{\rm smpl}}$
and  $  \langle \! \langle  N^{\rm E}_b(L) \rangle \! \rangle _{{\rm smpl}}$ become 
constants for large $L$ in each representation and these constants 
very weakly depend upon the representation. 

Results in Table I strongly indicate that the
restructuring is quite effective here. 
What is the reason for such qualitatively different behaviors in these
two representations then? To answer this question we plot in 
Fig.~2 the distributions of sorted coefficients in the expansion of 
$\mid  \psi_{\rm E}\rangle$ in both representations. 
The abscissa is the basis number $i'$ which is reordered so that $|c_{i'}|
\geq |c_{j'}|$ holds for $i' < j'$. 
We observe the
distributions are different in the region of the small
coefficients, where the normal representation has a longer tail.
The minimum value of the non-zero coefficients in the normal
representation is $7.6 \times 10^{-6}$ while it is $1.5 \times 10^{-4}$
in the restructured representation. We will return to this issue in an 
Appendix.
 
\section{Numerical results on a 8$\times$8 lattice}
 
Let us proceed to a larger lattice with 8$\times$8 sites and show that 
the SSS method is helpful to evaluate the energy of the first excited
states. We study the system with the total spin $S = 0$ and 1 
in the parameter region $0.55 \leq J'/J \leq 0.8$.
Here we use 
 \begin{eqnarray}
  E_{\{\eta\}}(L) = \langle \psi \mid  \hat H
M_{\{ \eta ^{(L)}\}} \hat H M_{\{ \eta ^{(L-1)}\}} \cdots \hat H
M_{\{ \eta ^{(1)}\}} \mid  \psi \rangle  
\end{eqnarray}
instead of (\ref{eldef}) in order to decrease the statistical errors.
Throughout this section we employ the restructured dimer representation,
which is a very powerful technique for the model under investigation.
The results we present here are compared to those in 
ref.17, where we have been successful to 
obtain approximate values on the energy difference for this system 
by means of the operator variational (OV) method.    

For this lattice size it is necessary to find an approximate state 
$\mid \! \psi_{\rm A} \rangle$ without knowing the exact eigenstate 
$\mid \! \psi_{\rm E} \rangle$. 
Here we employ a systematic way using the Lanczos-like method 
in small vector spaces; we start with a trial state 
$ \mid  \psi_{\rm trl} \rangle$, calculate the 
matrix elements of $\hat H$ from the basis 
$\{ \ \mid \! \psi_{\rm trl} \rangle, \hat H \! \mid  \psi_{\rm trl} \rangle,
\hat H^2 \! \mid  \psi_{\rm trl} \rangle, \cdots, 
\hat H^p \! \mid  \psi_{\rm trl} \rangle \}$
with a small integer $p$, numerically diagonalize the matrix to obtain 
its all energy eigenvalues 
and define $\mid \! \psi _{\rm A} \rangle$ as the
eigenstate which provides the lowest eigenvalue among them. 
In the following calculations we fix $p=6$.

As for the trial state $ \mid  \! \psi_{\rm trl} \rangle$ we employ one of the
states described in the appendix D in ref.17 whose energy
eigenvalue is found to be the lowest of all states we tried
in the numerical study by means of the OV method.    
Namely we start from 
$ \mid \! \psi_{\rm trl} \rangle$ = $\mid \! \Psi_{nn,0,(0,0),b'} \rangle
$ for $S=0$ and 
$ \mid \! \psi_{\rm trl} \rangle$ = $\mid \! \Psi_{nn,1,(0,0),d'} \rangle$ for
$S=1$,
which are among the two-nearest-neighboring-triplet states 
with the $odd$ and the $even$ $I_+$ parity, respectively.  
It should be noted that in our previous study\cite{ov}  
the energy eigenvalue for the $even$ $I_+$ is much higher
than that for the $odd$ $I_+$ when $S=0$ while for $S=1$ the lowest
energy is independent of the $I_+$ parity.   

Now we present the results. Table III and IV show values of  
$ \ \langle \! \langle E_{{\rm A}{\{\eta \}}}(L) \rangle \! \rangle _{{\rm smpl}} \ $
observed for the $S=0$ and $S=1$ cases, where the suffix A represents
that $\mid  \psi \rangle$ in (\ref{eldef}) is an approximate state 
$\mid \psi_{\rm A} \rangle$.
In both tables we see very small errors, which support that 
the restructuring technique is quite effective.
How many non-zero coefficients appear here?
When $S=0$ the number of non-zero coefficients in 
$\mid  \psi_{\rm A} \rangle$ is 
1,264,256, while
$\langle \! \langle  N^{\rm A}_b(5) \rangle \! \rangle _{{\rm smpl}} $, the    
number of non-zero coefficients in 
$\hat H M_{\{ \eta ^{(4)}\}} \hat H M_{\{ \eta ^{(3)}\}}
\hat H M_{\{ \eta ^{(2)}\}} \hat H M_{\{ \eta ^{(1)}\}} 
\mid  \psi_{\rm A} \rangle $ in measurements, ranges from
$(9.78 \pm 0.03) \times 10^5$ at $J'/J=0.55$ 
to $(1.912 \pm 0.004) \times 10^6$ at $J'/J=0.75$.
In the $S=1$ case $\mid  \psi_{\rm A} \rangle$ contains 690,624
components and we observe
$\langle \! \langle  N^{\rm A}_b(5) \rangle \! \rangle _{{\rm smpl}} $ is
$(2.84 \pm 0.02) \times 10^5$ ($(1.534 \pm 0.004) \times  10^6$) at 
$J'/J=0.55$ (0.8).
By operating $ M_{\{ \eta ^{(5)}\}}$ we can reduce it to less than a tenth.
In concrete terms, $\langle \! \langle  N^{\rm A}_a(5) \rangle \! \rangle _{{\rm smpl}} $
for $S=0$ ranges from $(8.35 \pm 0.02) \times 10^4 $ at $J'/J=0.55$ 
to $(1.657 \pm 0.003) \times 10^5$ at $J'/J=0.75$ and for $S=1$ it is 
$(2.27 \pm 0.01) \times 10^4$ ($(1.271 \pm 0.003) \times 10^5$) at 
$J'/J=0.55$ (0.8).
Thus we see that we can evaluate expectation values of 
$\hat H ^L $ not only for small $J'/J$ but also 
in the critical region of the parameter space. 
  
How do we extract energy eigenvalues from these data? We try two ways, 
one is to use the extrapolation formula and another is to do the 
Lanczos-like calculation with the basis 
$\{ \ \mid \! \psi_{\rm A} \rangle$, $\hat H \! \mid  \! \psi_{\rm A} \rangle$,
$\hat H^2 \! \mid \! \psi_{\rm A} \rangle \}$, noting that the
Hamiltonian matrix elements from this basis are given by 
$\langle \psi_{\rm A} \mid  \hat H^L \mid  \psi_{\rm A} \rangle$
up to $L=5$. 

The extrapolation formula is the one discussed in the previous paper\cite{mm}, 
\begin{eqnarray}
\langle \psi_{\rm A} \mid  \hat H^L \mid  \psi_{\rm A} \rangle  =
E^L( q_0 +\frac{q_1}{L+ \alpha +1})    \equiv   F(L,E,q_0,q_1,\alpha) \ .
\label{fitf}
\end{eqnarray}
In process of determining four parameters in this formula, $E$, $q_0$, $q_1$
and $\alpha$, we use two relations 
\begin{eqnarray*}
F(0,E,q_0,q_1,\alpha) & = & q_0 +\frac{q_1}{\alpha +1} = 
\langle \psi_{\rm A} \mid \psi_{\rm A} \rangle = 1 \ , \\
F(1,E,q_0,q_1,\alpha) & = & E \cdot \left( q_0 +\frac{q_1}{\alpha +2} \right) =
\langle \psi_{\rm A} \mid  \hat H \mid  \psi_{\rm A} \rangle \ , 
\end{eqnarray*}
so that $q_1$ and $\alpha$ are calculated from $E$ and $q_0$.
Note that the value of 
$\langle \psi_{\rm A} \mid   \hat H \mid  \psi_{\rm A} \rangle$ is known 
in the process of generating $\mid  \psi_{\rm A} \rangle$. 
Then we look for the values of $E$ and $q_0$ which minimize 
\begin{eqnarray}
 S_{\rm fit} \equiv \sum_{L=1}^{5} \left[1-
\frac{ \ \langle \! \langle E_{{\rm A}{\{\eta \}}}(L) \rangle \!
\rangle _{{\rm smpl}}}{F(L,E,q_0,q_1,\alpha)}\right] ^2 \  ,
\end{eqnarray}
within the ranges 
$\langle \psi_{\rm A} \mid   \hat H \mid  \psi_{\rm A} \rangle - 0.5 \leq
E \leq \langle \psi_{\rm A} \mid   \hat H \mid  \psi_{\rm A} \rangle$ and
$0.5 \leq q_0  \leq 1.0$. 
Results of the fitting for $J'/J > 0.5$ 
are shown in Tables V ($S=0$) and VI ($S=1$). 
The statistical errors of the measured values, 
\begin{eqnarray}
S_{\rm 2Er} \equiv \sum_{L=1}^{5} \left[
\frac{Er(L)}{ \ \langle \! \langle E_{{\rm A}{\{\eta \}}}(L) \rangle \!
\rangle _{{\rm smpl}}}\right] ^2 \ ,
\end{eqnarray}
are also presented in the tables. The fact that $ S_{\rm fit}$ is smaller than
$S_{\rm 2Er}$ means the fitting is statistically acceptable. 
We see that values of $q_0$ obtained from the fit are close to 1. Since 
$q_0$ represents the overlapping between the trial state and the exact
state, $\langle \psi_{\rm A} \mid  \psi_{\rm E} \rangle$, this implies  
that our way to generate the state $\mid  \psi_{\rm A} \rangle$ supplies 
good approximations of the exact eigenstate $\mid  \psi_{\rm E} \rangle$
for this model.
The energy difference from the fit normalized by the intra-dimer coupling, 
$(E_{\rm fit} - E_0)/J$, is plotted in Fig.~3 ($S=0$) and Fig.~4 $(S=1)$ 
together with those obtained by the OV method. Here $E_0$ denotes the
energy of the ground state in the singlet dimer phase, which is $-24.0J$
for the $8 \times 8$ system. We see that the data 
from the OV method with $L_{max} = 4$ and the data from the SSS method 
are in good agreement when $J'/J = 0.55$. For larger coupling ratios 
the data from the SSS method are below the data from the OV method. 
Especially near the critical point this tendency is outstanding, 
suggesting that the SSS method is more suitable to study the phase
transition in the SS model. The data by the SSS method also support that
an intermediate spin-gap phase exists, 
because at $J'/J \simeq 0.71$ $E_{\rm fit} - E_0 $ becomes zero 
for $S=0$ while $E_{\rm fit}-E_0 \sim 0.18 J$ for $S=1$ there. 

Let us turn to another way we try to extract energy eigenvalues. 
Based on the observation that $\mid  \psi_{\rm A} \rangle$ is a good 
approximate of $\mid  \psi_{\rm E} \rangle$, we apply the Lanczos-like method 
to evaluate the eigenvalue $E^{(n)}$ ($n$= 1, 2, 3) from 
the matrix $H^{(n)}$ whose matrix elements are calculated from bases 
$\{ \mid \! \psi_{\rm A} \rangle \}$,
$\{  \mid  \!\psi_{\rm A} \rangle \ , 
\hat H \!\mid \! \psi_{\rm A} \rangle \}$ and  
$\{  \mid \! \psi_{\rm A} \rangle \ , 
\hat H \!\mid  \!\psi_{\rm A} \rangle \ , 
\hat H^2 \!\mid \! \psi_{\rm A} \rangle \}$, respectively. 
Substituting 
$\langle \! \langle E_{{\rm A}{\{\eta \}}}(L) \rangle \! \rangle _{{\rm smpl}}$
in Tables III and IV for
$E_{\rm A}(L) = \langle \psi_{\rm A} \! \mid  \!\hat H^L  \! \mid  \!
\psi_{\rm A}\rangle $ in 
\begin{eqnarray*}
\alpha_1 &\equiv& \langle \psi_{\rm A} \mid  \hat H \mid  \psi_{\rm A}\rangle  \ , \\
\beta_1 &\equiv& \sqrt{\langle \psi_{\rm A} \mid  \hat H^2 \mid  \psi_{\rm A}\rangle
- \alpha_1 ^2} \ , \\
\alpha_2 &\equiv&  \frac{1}{\beta_1^2} 
\left( \langle \psi_{\rm A} \mid  \hat H^3 \mid  \psi_{\rm A}\rangle
- \alpha_1 ^3 \right) -2 \alpha_1  \ ,\\
\beta_2 &\equiv& \sqrt{\frac{1}{\beta_1 ^2}
\left( \langle \psi_{\rm A} \mid  \hat H^4 \mid  \psi_{\rm A}\rangle
- 2 \alpha_1 \langle \psi_{\rm A} \mid  \hat H^3 \mid  \psi_{\rm A}\rangle
+ \alpha_1 ^2  \langle \psi_{\rm A} \mid  \hat H^2 \mid  \psi_{\rm A}\rangle
\right)-\alpha_2 ^2 -\beta_1 ^2}  \ , \\
\alpha_3 &\equiv& \frac{1}{\beta_1^2 \beta_2^2}
\left( \langle \psi_{\rm A} \mid  \hat H^5 \mid  \psi_{\rm A}\rangle
- 2 \alpha_1 \langle \psi_{\rm A} \mid  \hat H^4 \mid  \psi_{\rm A}\rangle
+ \alpha_1 ^2  \langle \psi_{\rm A} \mid  \hat H^3 \mid  \psi_{\rm A}\rangle 
\right) -2 \alpha_2 \\
&-&\frac{\beta_1^2}{\beta_2^2}\left( \alpha_1 +2 \alpha_2 \right) 
-\frac{\alpha_2 ^3}{\beta_2^2}  \ ,
\end{eqnarray*}
we diagonalize 
\begin{eqnarray*}
H^{(1)} \equiv \left( \alpha_1 \right)  , \ \ \
H^{(2)} \equiv \left( \begin{array}{cc} \alpha_1 & \beta_1 \\
\beta_1 & \alpha_2 \end{array} \right) , \ \ \ 
H^{(3)} \equiv \left( \begin{array}{ccc}  \alpha_1 & \beta_1 & 0 \\
\beta_1 & \alpha_2 & \beta_2 \\ 0 & \beta_2 & \alpha_3 \end{array} \right)  . 
\end{eqnarray*}
The results are also in Tables V ($S=0$) and VI ($S=1$). As we see in
tables we obtain better estimates by $E^{(2)}$ than by $E^{(1)}$. 
Note that values of $E^{(2)}$ are in good accordance with $E_{\rm fit}$ 
in Tables III and IV, which indicates that the fits are reliable.
We could not, on the other hand, obtain trustworthy values of 
$E^{(3)}$ from the present data. It is mainly  
because of the insufficient statistical precision, which we can cope 
with by increasing the number of samples, or by decreasing the parameter 
$\epsilon$. It seems that the latter is much more effective than the former.
In the $S=0$ and $J'/J=0.7$ case, for instance, we observe that we
cannot obtain any $E^{(3)}$ from $9 \times 10^4$ samples with 
$\epsilon = 0.001$ while $E^{(3)}$ calculated from $10^4$ samples with 
$\epsilon = 0.0005$ is $-23.928 J$, which is slightly lower than the 
value of $E^{(2)}$ in Table V. 

\section{Summary and Discussions}

In this paper we calculated the expectation values of $\hat H ^L$
$(L=1,2, \cdots)$ of the two-dimensional SS model\cite{ss} using 
the SSS method\cite{mm}.  
Comparing our results on a $4 \times 4$ lattice with exact values we
demonstrated that the method is applicable even to 
this strongly frustrated quantum spin system. 
We also showed that deviations of the evaluated values are much less
in the restructured representation than in the normal representation.  
In the normal representation of the SS model we observed rapid increase
of the statistical errors for large values of $L$, 
which is in contrast with our previous study of the $4 \times 4 $ spin 
one-half Heisenberg model\cite{mm}.  
A qualitative discussion on the reason why such increase occurs is given
in the appendix. 

In study of the $8 \times 8$ SS system we concentrated our attention on 
an intermediate phase reported in refs.11-17. 
We examined energy
eigenvalues of the lowest excited states with the total spin $S=0$ and
$S=1$ for several values of the coupling ratio.   
It should be emphasized that the method is powerful even in the critical
region of the model, where the negative sign 
problem prevents us from obtaining statistically meaningful results 
by means of the standard quantum Monte Carlo techniques\cite{foot1}. 
In order to numerically extract the energy eigenvalues from the
observed $\langle \hat H ^L \rangle$ we tried two ways, a fitting within
some parameter space and a Lanczos-like diagonalization. We see results 
from these two ways are consistent to give us reliable estimate of
energy eigenvalues. It should be kept in mind that employing highly qualified
trial states in the SSS method is a key for the success. 
Thus we see that the critical value of $J'/J$ is less than 0.71 
for the phase transition from the singlet dimer phase to the
intermediate phase, which is compatible with the value in
ref.12. Note that this upper bound is obtained with a 
non-perturbative method. The spin gap at the critical point, 
$ \sim 0.18 J$, is about one-half of the one
reported in ref.12, however. More intensive investigations should be
made in future work before we establish physical properties of this phase.
  
\appendix
\section{Large Variances}
In this appendix we discuss the reason why very large variances are observed 
in the normal representation of the SS model. 
As we have seen in Fig.~2 much more terms with very small
coefficients appear in the expansion of the wave function $\mid \psi_{\rm A}
\rangle$ in the normal representation compared to the restructured 
representation. So we analyze the expression for the variance paying 
our attention to contributions from those terms. 

Let us examine $\sigma^2_{\{\eta \}}(L)$, the variance of 
$\langle \! \langle E_{\{\eta \}}(L) \rangle \! \rangle$ introduced 
in ref. 4,
\begin{eqnarray}
\sigma_{\{\eta \}}^2(L) &\equiv&
\langle \! \langle E_{\{\eta \}}(L)^2 \rangle \! \rangle -
\langle \! \langle E_{\{\eta \}}(L) \rangle \! \rangle \ ^2 \nonumber \\
&=&  \sum_{i,i'}  \sum_{j,j'} \cdots \sum_{k,k'} \sum_{l,l'}
c_i h_{ij}\cdots h_{kl}c_l \cdot c_{i'} h_{i'j'}\cdots h_{k'l'} c_{l'}
 \nonumber \\
& & \times \ \langle \! \langle  \eta_i^{(1)}\eta_{i'}^{(1)}
\rangle \! \rangle \langle \! \langle  \eta_j^{(2)}\eta_{j'}^{(2)}
\rangle \! \rangle   \cdots
\langle \! \langle  \eta_k^{(L)}\eta_{k'}^{(L)}
\rangle \! \rangle \langle \! \langle  \eta_l^{(L+1)}\eta_{l'}^{(L+1)}
\rangle \! \rangle \ - E(L)^2
\nonumber \\
&=&  \sum_{i,i'}  \sum_{j,j'} \cdots \sum_{k,k'} \sum_{l,l'}
c_i h_{ij}\cdots h_{kl}c_l \cdot c_{i'} h_{i'j'}\cdots h_{k'l'} c_{l'}
\cdot \{1 + \delta_{ii'}
(  \langle \! \langle  \eta_i^2 \rangle \! \rangle -1) \}
\nonumber \\
& & \times \
\{1 + \delta_{jj'}(\langle \! \langle  \eta_j^2 \rangle \! \rangle -1)\}
 \cdots
\{1 + \delta_{kk'}( \langle \! \langle  \eta_k^2 \rangle \! \rangle -1)\}
\{1 + \delta_{ll'}( \langle \! \langle  \eta_l^2 \rangle \! \rangle -1)\} 
\nonumber \\
& -&  E(L)^2 \ .
\label{s2etal}
\end{eqnarray}
For $L \geq 2$ this variance contains one or more terms that
might blow up with small non-zero coefficients,  
because the statistical average 
$\langle \! \langle   \eta^2 \rangle \! \rangle
= \epsilon / |c_s| $, which follows from (\ref{eta2}) and (\ref{ai}), 
becomes huge if a non-zero coefficient $c_s$ in (\ref{pi}) is much less than 
$\epsilon$ . 
The most dangerous term among them is
\begin{eqnarray*}
T &\equiv& \sum_{i,j,k,\cdots,l,m,n}  c_i^2 c_n^2
(h_{ij})^2 (h_{jk})^2 \cdots (h_{lm})^2 (h_{mn})^2 \\
&\times&
(\langle \! \langle   \eta_i^2 \rangle \! \rangle -1)
(\langle \! \langle   \eta_j^2 \rangle \! \rangle -1)
(\langle \! \langle   \eta_k^2 \rangle \! \rangle -1)
\cdots
(\langle \! \langle   \eta_l^2 \rangle \! \rangle -1)
(\langle \! \langle   \eta_m^2 \rangle \! \rangle -1)
(\langle \! \langle   \eta_n^2 \rangle \! \rangle -1). 
\end{eqnarray*} 
For $L=1$ this term is harmless since in this case 
\begin{eqnarray*}
T &=& \sum_i c_i^2
(\langle \! \langle   \eta_i^2 \rangle \! \rangle -1) \sum_j c_j^2 
(\langle \! \langle   \eta_j^2 \rangle \! \rangle -1) (h_{ij})^2 \\
&=& \sum_{c_i \neq 0} c_i^2 \left[
{\rm max}\left( 1,\frac{\epsilon}{|c_i|} \right)-1 \right]
\sum_{c_j \neq 0} c_j^2 \left[
{\rm max}\left( 1,\frac{\epsilon}{|c_j|} \right)-1 \right](h_{ij})^2 \\
&=& \sum_{0 < |c_i| < \epsilon}  (\epsilon |c_i|  - c_i^2)  
\sum_{0 < |c_j| < \epsilon} (\epsilon |c_j|  - c_j^2)(h_{ij})^2 \ .
\end{eqnarray*} 
Actually we observe on a $4\times 4 $ lattice that the measured variances of 
$\langle \! \langle E_{\{\eta \}}(1) \rangle \! \rangle$, which are
defined by (\ref{wem5}), are comparable in both representations. 
They are $0.153 J^2$ and $0.135 J^2$ when $\epsilon = 0.1$ and
$n_{{\rm smpl}}= 10^4$ for the normal representation and the restructured 
representation, respectively.    

Let us then consider the case $L=2$, where
\begin{eqnarray*}
T &=& \sum_i c_i^2 
(\langle \! \langle   \eta_i^2 \rangle \! \rangle -1) \sum_k c_k^2 
(\langle \! \langle   \eta_k^2 \rangle \! \rangle -1)
\sum_j (h_{ij})^2 (h_{jk})^2
(\langle \! \langle   \eta_j^2 \rangle \! \rangle -1) \\
&=& \sum_{0 < |c_i| < \epsilon} (\epsilon |c_i|  - c_i^2) 
\sum_{0 < |c_k| < \epsilon}(\epsilon |c_k|  - c_k^2)
\sum_j (h_{ij})^2 (h_{jk})^2
(\langle \! \langle   \eta_j^2 \rangle \! \rangle -1) \\
&\sim& \sum_{0 < |c_i| < \epsilon} (\epsilon |c_i|  - c_i^2) 
\sum_{0 < |c_k| < \epsilon}(\epsilon |c_k|  - c_k^2)
\sum_{0 < |c_j|  \ll \epsilon}(h_{ij})^2 (h_{jk})^2 \frac{\epsilon}{|c_j| } \ .
\end{eqnarray*}
If the Hamiltonian matrix is positive-definite ($h_{jk} \geq 0 $) and
all coefficients of the eigen state are non-negative ($c_k \geq 0$), 
\begin{eqnarray*}
0 \leq h_{jk}c_k \leq \sum_l h_{jl}c_l = E c_j 
\end{eqnarray*}
holds for $any$ $k$. In this case it follows that
\begin{eqnarray*} 
T &\sim& \sum_{0 < c_i < \epsilon} (\epsilon c_i  - c_i^2) 
\sum_{0 < c_k < \epsilon}(\epsilon  - c_k)
\sum_{0 < c_j  \ll \epsilon}(h_{ij})^2 h_{jk}\cdot h_{jk}c_k \cdot 
\frac{\epsilon}{c_j} \\  
&\leq& \sum_{0 < c_i< \epsilon}(\epsilon c_i - c_i^2) 
\sum_{0 < c_k< \epsilon} (\epsilon - c_k) \sum_{0 < c_j \ll \epsilon}
(h_{ij})^2 h_{jk} \cdot  Ec_j \cdot  \frac{\epsilon}{c_j} \\ 
&=&\epsilon E \sum_{0 < c_i< \epsilon}(\epsilon c_i - c_i^2) 
\sum_{0 < c_k< \epsilon}(\epsilon - c_k)  \sum_{0 < c_j \ll \epsilon}
(h_{ij})^2 h_{jk}  \ ,
\end{eqnarray*}
which gives a safe upper bound of $T$. This means that the 
positive-definite systems are free from unacceptably large variances 
caused by the smallness of the coefficients. 
For the frustrated systems with the negative sign problem, on the contrary, 
there is no guarantee that they can escape from this difficulty. 
It is very likely, therefore, that the frustration and a lot of small
coefficients would be responsible together
to such behavior of the statistical errors in the measurement.  
Clearly similar discussions are possible for $L \geq 3$. 


\newpage

\begin{table}[ht]
\centering
\begin{tabular}{rrlrlrl} \hline
\multicolumn{1}{c}{ } &  \multicolumn{2}{c}{ }
&\multicolumn{2}{c}{Normal}
  &\multicolumn{2}{c}{Restructured}
\\ \cline{4-7}
\multicolumn{1}{c}{$L$}
& \multicolumn{2}{c}{$E_{\rm E}(L)/J^L$} & \multicolumn{4}{c}
{$\langle \! \langle E_{{\rm E}{\{\eta \}}}(L) \rangle \! \rangle _{{\rm smpl}}/J^L$}
 \\ \hline
  1  &  $-$0.63502 &$ \times  10^{ 1}$ & $-$(0.63484  $\pm $ 0.00078) &
$ \times  10^{ 1}$ & $-$(0.63486  $\pm $ 0.00073) &
$ \times  10^{ 1}$ \\ \hline
  2  &  0.40325 & $ \times  10^{ 2}$ & (0.40336 $\pm $ 0.00071) &
 $ \times  10^{ 2}$ & (0.40298 $\pm $ 0.00063)&
 $ \times  10^{ 2}$ \\ \hline
  3  & $-$0.25607 & $ \times  10^{ 3}$ & $-$(0.25580 $\pm $ 0.00096) &
$ \times  10^{ 3}$ & $-$(0.25601 $\pm $ 0.00052)&
 $ \times  10^{ 3}$ \\ \hline
  4  &  0.16261 & $ \times  10^{ 4}$ & (0.1619 $\pm $ 0.0023)&
 $ \times  10^{ 4}$ & (0.16277  $\pm $ 0.00052)&
 $ \times  10^{ 4}$ \\ \hline
  5  & $-$0.10326 & $ \times  10^{ 5}$ & $-$(0.1056 $\pm $ 0.0052)&
 $ \times  10^{ 5}$ & $-$(0.10307 $\pm $ 0.00053)&
 $ \times  10^{ 5}$ \\ \hline
  6  &  0.65576 & $ \times  10^{ 5}$ & (0.692$\pm $ 0.043) &
 $ \times  10^{ 5}$ & (0.6568 $\pm $ 0.0042) &
 $ \times  10^{ 5}$ \\ \hline
  7  & $-$0.41642 & $ \times  10^{ 6}$ & $-$(0.425 $\pm $ 0.053)&
 $ \times  10^{ 6}$ & $-$(0.4148 $\pm $ 0.0035) &
$ \times  10^{ 6}$ \\ \hline
  8  &  0.26443 & $ \times  10^{ 7}$ & (0.277 $\pm $ 0.081)&
 $ \times  10^{ 7}$ & (0.2655 $\pm $ 0.0034) &
$ \times  10^{ 7}$ \\ \hline
  9  & $-$0.16792 & $ \times  10^{8}$ & $-$(0.06 $\pm $ 0.19) &
$ \times  10^{8}$ & $-$(0.1701 $\pm $ 0.0025)&
 $ \times  10^{8}$ \\ \hline
 10  &  0.10663 & $ \times  10^{9}$ & (0.107 $\pm $ 0.048) &
$ \times  10^{9}$ & (0.1048 $\pm $ 0.0030) & $ \times  10^{9}$ \\ \hline
\end{tabular}
\caption{
Results on
$\langle \! \langle E_{{\rm E}{\{\eta \}}}(L) \rangle \! \rangle _{{\rm smpl}}$
$(L=1,2, \cdots, 10)$ obtained for the $4 \times 4$ 
Shastry-Sutherland (SS) model by the 
Statistic State Selection (SSS) method. The number of samples is  
$10^4$ and $\epsilon = 0.1$ for both of the normal and the restructured
representations. 
Exact values of $E_{\rm E}(L)$ are also presented for comparison.}
\end{table}

\begin{table}[hb]
\centering
\begin{tabular}{rcccc} \hline
 { }&\multicolumn{2}{c}{Normal}
  &\multicolumn{2}{c}{Restructured}   \\ \cline{2-5}
\multicolumn{1}{c}{$L$}
&$ \ \langle \! \langle  N^{\rm E}_b(L) \rangle \! \rangle _{{\rm smpl}}$   & $ \ \langle \!
 \langle  N^{\rm E}_a(L) \rangle \! \rangle _{{\rm smpl}} $  &
$ \ \langle \! \langle  N^{\rm E}_b(L) \rangle \! \rangle _{{\rm smpl}}$ &
 $ \ \langle \! \langle  N^{\rm E}_a(L) \rangle \! \rangle _{{\rm smpl}} $  \\ \hline
 1 &  12870  &  577.7 &    8565 &     641.0 \\ \hline
 2&  7250.2 &    486.5&    7464.7 &   532.8 \\ \hline
 3 &  6353.7&   468.0&  6494.0&    502.8 \\ \hline
4 &   6159.0 &   463.3&  6189.0 &   492.8 \\ \hline
5&    6110.6 &   462.4&    6079.7 &   488.9 \\ \hline
6 &   6101.0 &   462.2&    6037.5 &   487.5 \\ \hline
7&    6097.7 &   462.3&    6020.0 &   487.0 \\ \hline
8 &   6099.1 &    461.9 &  6015.6 &   486.6 \\ \hline
9 &   6097.0 & 462.0&      6010.9 &   486.5 \\ \hline
10 &  6097.7 & 462.0&      6009.5 &   486.3 \\ \hline
\end{tabular}
\caption{
Numbers of non-zero coefficients $before$ and $after$ operating the
random choice matrix $M_{\{ \eta ^{(L)}\}}$ to the state
$\hat H M_{\{ \eta ^{(L-1)}\}} \cdots \hat H
M_{\{ \eta ^{(1)}\}} \mid  \psi_{\rm E} \rangle $
for the $4 \times 4$ SS model 
obtained from $10^4$ samples with $\epsilon = 0.1$.
}
\end{table}

\begin{table}[ht]
\centering
\small
\begin{tabular}{rrlrl} \hline
\multicolumn{1}{c}{$L$} &\multicolumn{4}{c}
{$\langle \! \langle E_{{\rm A}{\{\eta \}}}(L) \rangle \! \rangle _{{\rm smpl}}/J^L$}
\\ \cline{2-5}

\multicolumn{1}{c}{ }
& \multicolumn{2}{c}{$J'/J=0.55$} & \multicolumn{2}{c}{$J'/J=0.60$}
 \\ \hline
  1  &$-$(0.23268865 $\pm $ 0.00000027) &$ \times  10^{ 2}$ 
    & $-$(0.23467222 $\pm $ 0.00000030) &$ \times  10^{ 2}$ \\ \hline
  2  & (0.5414619 $\pm $ 0.0000011) &$ \times  10^{ 3}$ 
     & (0.5507468 $\pm $ 0.0000012)&$ \times  10^{ 3}$ \\ \hline
  3  & $-$(0.12600061 $ \pm $ 0.00000047) &$ \times  10^{ 5}$ 
     & $-$(0.12925980 $\pm $ 0.00000050)&$ \times  10^{ 5}$ \\ \hline
  4  & (0.2932180 $\pm $ 0.0000036)& $ \times  10^{ 6}$ 
     & (0.3033880 $\pm $ 0.0000032)&$ \times  10^{ 6}$ \\ \hline
  5  & $-$(0.682344 $\pm $ 0.000021)&$ \times  10^{ 7}$ 
     & $-$(0.712097 $\pm $ 0.000032)&$ \times  10^{ 7}$ 
\\ \hline

\multicolumn{1}{c}{ }
& \multicolumn{2}{c}{$J'/J=0.65$} & \multicolumn{2}{c}{$J'/J=0.70$}
 \\ \hline
  1  &$-$(0.23680656 $\pm $ 0.00000033) &$ \times  10^{ 2}$ 
    & $-$(0.23908546 $\pm $ 0.00000036) &$ \times  10^{ 2}$ \\ \hline
  2  & (0.5608338 $\pm $ 0.0000014) &$ \times  10^{ 3}$ 
     & (0.5717199 $\pm $ 0.0000015)&$ \times  10^{ 3}$ \\ \hline
  3  & $-$(0.13283432 $\pm $ 0.00000055) &$ \times  10^{ 5}$ 
     & $-$(0.13673295 $\pm $ 0.00000061)&$ \times  10^{ 5}$ \\ \hline
  4  & (0.3146405 $\pm $ 0.0000031)& $ \times  10^{ 6}$ 
     & (0.3270462 $\pm $ 0.0000032)&$ \times  10^{ 6}$ \\ \hline
  5  & $-$(0.745321 $\pm $ 0.000023)&$ \times  10^{ 7}$ 
     & $-$(0.782349 $\pm $ 0.000030)& $ \times  10^{ 7}$ 
\\ \hline 

\multicolumn{1}{c}{ }
& \multicolumn{2}{c}{$J'/J=0.75$} & \multicolumn{2}{c}{ }
 \\ \hline
  1  &$-$(0.24149807 $\pm $ 0.00000039) &$ \times  10^{ 2}$ 
&　 &　\\ \hline
  2  & (0.5833748 $\pm $ 0.0000016) &$ \times  10^{ 3}$ 
&　 &　\\ \hline
  3  & $-$(0.14095261 $\pm$ 0.00000067) &$ \times  10^{ 5}$ 
&　 &　\\ \hline
  4  & (0.3406212 $\pm $ 0.0000035)&$ \times  10^{ 6}$ 
&　 &　\\ \hline
  5  & $-$(0.823245 $\pm $ 0.000028)& $ \times  10^{ 7}$ 
& 　 &　\\ \hline 
\end{tabular}
\caption{
$S=0$ results on 
$\langle \! \langle E_{{\rm A}{\{\eta \}}}(L) \rangle \! \rangle _{{\rm smpl}}$
$(L=1,2, \cdots, 5)$ obtained for the $8 \times 8$ SS model by the 
SSS method
with $n_{{\rm smpl}} = 10^4$, $\epsilon = 0.001$ and $\delta = 0.00001$.}
\end{table}

\begin{table}[ht]
\centering
\small
\begin{tabular}{rrlrl} \hline
\multicolumn{1}{c}{$L$} &\multicolumn{4}{c}
{$\langle \! \langle E_{{{\rm A}\{\eta \}}}(L) \rangle \! \rangle _{{\rm smpl}}/J^L$}
\\ \cline{2-5}

\multicolumn{1}{c}{ }
& \multicolumn{2}{c}{$J'/J=0.55$} & \multicolumn{2}{c}{$J'/J=0.60$}
 \\ \hline
  1  &$-$(0.23403813 $\pm $ 0.00000013) &$ \times  10^{ 2}$ 
    & $-$(0.23504307 $\pm $ 0.00000016) &$ \times  10^{ 2}$ \\ \hline
  2  & (0.54774476 $\pm $ 0.00000056) &$ \times  10^{ 3}$ 
     & (0.55246007 $\pm $ 0.00000065)&$ \times  10^{ 3}$ \\ \hline
  3 & $-$(0.12819601 $\pm $ 0.00000026) &$ \times  10^{ 5}$ 
    & $-$(0.12985518 $\pm $ 0.00000029)&$ \times  10^{ 5}$ \\ \hline
  4  & (0.3000356 $\pm $ 0.0000021)&$ \times  10^{ 6}$ 
     & (0.3052268 $\pm $ 0.0000022)&$ \times  10^{ 6}$ \\ \hline
  5  & $-$(0.702218 $\pm $ 0.000023)& $ \times  10^{ 7}$ 
     & $-$(0.717444 $\pm $ 0.000027)&$ \times  10^{ 7}$ 
\\ \hline 

\multicolumn{1}{c}{ }
& \multicolumn{2}{c}{$J'/J=0.65$} & \multicolumn{2}{c}{$J'/J=0.70$}
 \\ \hline
  1  &$-$(0.23622014 $\pm $ 0.00000019) &$ \times  10^{ 2}$ 
    & $-$(0.23762394 $\pm $ 0.00000024) &$ \times  10^{ 2}$ \\ \hline
  2  & (0.55801950 $\pm $ 0.00000081) &$ \times  10^{ 3}$ 
     & (0.5647149 $\pm $ 0.0000010)&$ \times  10^{ 3}$ \\ \hline
  3  & $-$(0.13182375 $\pm $ 0.00000034) &$ \times  10^{ 5}$ 
     & $-$(0.13421698 $\pm $ 0.00000042)&$ \times  10^{ 5}$ \\ \hline
  4  & (0.3114219 $\pm $ 0.0000023)& $ \times  10^{ 6}$ 
     & (0.3190188 $\pm $ 0.0000023)&$ \times  10^{ 6}$ \\ \hline
  5  & $-$(0.735702 $\pm $ 0.000014)&$ \times  10^{ 7}$ 
     & $-$(0.758308 $\pm $ 0.000017)& $ \times  10^{ 7}$ 
\\ \hline 

\multicolumn{1}{c}{ }
& \multicolumn{2}{c}{$J'/J=0.75$} & \multicolumn{2}{c}{$J'/J=0.80$ }
 \\ \hline
  1  &$-$(0.23936605 $\pm $ 0.00000029) &$ \times  10^{ 2}$ 
    & $-$(0.24150001 $\pm $ 0.00000036) &$ \times  10^{ 2}$ \\ \hline
  2  & (0.5731340 $\pm $ 0.0000012) & $ \times  10^{ 3}$ 
     & (0.5835473 $\pm $ 0.0000015)&$ \times  10^{ 3}$ \\ \hline
  3  & $-$(0.13726176 $\pm $ 0.00000049) &$ \times  10^{ 5}$ 
     & $-$(0.14106434 $\pm $ 0.00000057)&$ \times  10^{ 5}$ \\ \hline
  4  & (0.3287936 $\pm $ 0.0000028)& $ \times  10^{ 6}$ 
     & (0.3411166 $\pm $ 0.0000030)&$ \times  10^{ 6}$ \\ \hline
  5  & $-$(0.787688 $\pm $ 0.000021)& $ \times  10^{ 7}$ 
     & $-$(0.825086 $\pm $ 0.000022)&$ \times  10^{ 7}$ 
\\ \hline 
\end{tabular}
\caption{
$S=1$ results on 
$\langle \! \langle E_{{\rm A}{\{\eta \}}}(L) \rangle \! \rangle _{{\rm smpl}}$
$(L=1,2, \cdots, 5)$ obtained for the $8 \times 8$ SS model 
by the SSS method
with $n_{{\rm smpl}} = 10^4$, $\epsilon = 0.001$ and $\delta = 0.00001$.}
\end{table}

\begin{table}[ht]
\centering
\begin{tabular}{rrlrlccccc} \hline
\multicolumn{1}{c}{$J'/J$ } &  \multicolumn{2}{c}{$S_{\rm 2Er}$} &
\multicolumn{2}{c}{$S_{{\rm fit}}$ } & \multicolumn{1}{c}{$q_{0 \ {\rm fit}}$ } &
\multicolumn{1}{c}{$E_{{\rm fit}}/J$ } & \multicolumn{1}{c}{$E^{(1)}/J$} &
\multicolumn{1}{c}{$E^{(2)}/J$} & \multicolumn{1}{c}{$E^{(3)}/J$} 
\\ \hline   
0.55 & 1.1  &$ \times  10^{ -9}$ & 3.7 &$ \times  10^{-10}$ & 0.995 &
$-$23.276 & $-$23.269 & $-$23.272 & $-$23.273 \\ \hline
0.60 & 2.1  &$ \times  10^{ -9}$ & 8.4  &$ \times  10^{ -11}$ & 0.994 &
 $-$23.478 & $-$23.467 & $-$23.473 &  $-$23.476  \\ \hline
0.65 & 1.1 &$ \times  10^{ -9}$ & 8.5  &$ \times  10^{ -11}$ & 0.988 &
$-$23.700  & $-$23.681 &  $-$23.691 & $-$58.851 \\ \hline
0.70 & 1.6  &$ \times  10^{ -9}$ & 1.7  &$ \times  10^{ -10}$ & 0.979 &
 $-$23.942  & $-$23.909 & $-$23.927  & \multicolumn{1}{c}{$-$ }\\ \hline
0.75 & 1.3  &$ \times  10^{ -9}$ & 3.6  &$ \times  10^{ -10}$ & 0.975 &
$-$24.196 & $-$24.150 & $-$24.178 & $-$24.183 \\ \hline 
\end{tabular}
\caption{Results from the fit and from the Lanczos-like evaluation 
for S=0 data in Table III.}
\end{table}

\begin{table}[h]
\centering
\begin{tabular}{rrlrlccccc} \hline
\multicolumn{1}{c}{$J'/J$ } &  \multicolumn{2}{c}{$S_{\rm 2Er}$} &
\multicolumn{2}{c}{$S_{{\rm fit}}$ } & \multicolumn{1}{c}{$q_{0 \ {\rm fit}}$ } &
\multicolumn{1}{c}{$E_{{\rm fit}}/J$ } & \multicolumn{1}{c}{$E^{(1)}/J$} &
\multicolumn{1}{c}{$E^{(2)}/J$} & \multicolumn{1}{c}{$E^{(3)}/J$}
\\ \hline   
0.55 & 1.1  &$ \times  10^{ -9}$ & 4.9 &$ \times  10^{-11}$ & 0.986 &
$-$23.410 & $-$23.404 & $-$23.408 & \multicolumn{1}{c}{$-$ } \\ \hline
0.60 & 1.4  &$ \times  10^{ -9}$ & 5.4  &$ \times  10^{ -12}$ & 0.994 &
 $-$23.509 & $-$23.504 & $-$23.506 &  $-$23.507 \\ \hline
0.65 & 4.2 &$ \times  10^{ -10}$ & 2.7  &$ \times  10^{ -10}$ & 0.995 &
$-$23.629 & $-$23.622 &  $-$23.626 & $-$23.626\\ \hline
0.70 & 5.6  &$ \times  10^{ -10}$ & 4.9  &$ \times  10^{ -11}$ & 0.991 &
 $-$23.780 & $-$23.762 & $-$23.776  & \multicolumn{1}{c}{$-$ } \\ \hline
0.75 & 8.2  &$ \times  10^{ -10}$ & 2.2  &$ \times  10^{ -10}$ & 0.978 &
$-$23.982 & $-$23.937 & $-$23.967 & $-$23.969 \\ \hline
0.80 &8.1  &$ \times  10^{ -10}$ & 3.8  &$ \times  10^{ -10}$ & 0.971 &
$-$24.223 & $-$24.150 & $-$24.205 & $-$24.209 \\ \hline 
\end{tabular}
\caption{Results from the fit and from the Lanczos-like evaluation 
for S=1 data in Table IV.}
\end{table}

\begin{figure}[p]
\begin{center}
\scalebox{0.5}{\includegraphics{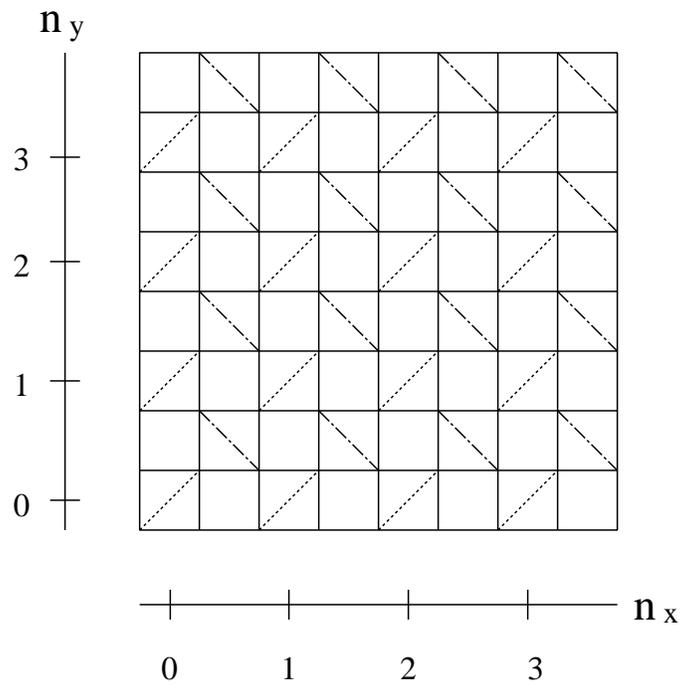}}
\caption{A schematic view of the Shastry-Sutherland (SS) 
model on a $8 \times 8$
lattice. Solid lines represent the inter-dimer coupling ($J'$) and other
 lines do the intra-dimer coupling ($J$).  
}
\end{center}
\end{figure}
\begin{figure}[b]
\begin{center}
\scalebox{0.5}{\includegraphics{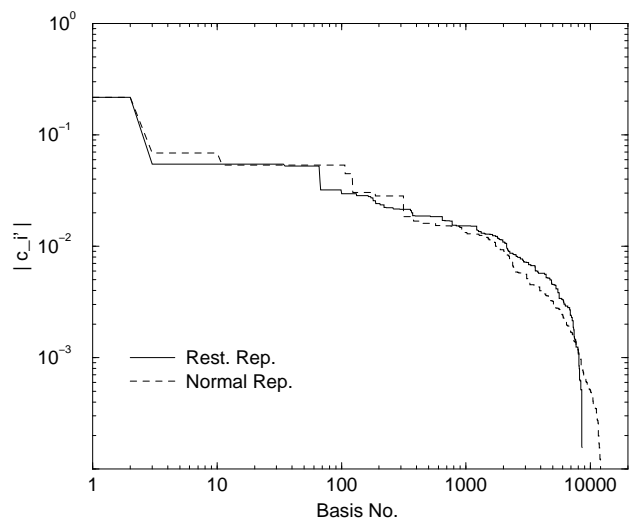}}
\caption{Distributions of coefficients for the $4 \times 4$ SS model 
in the normal representation (the dashed line) and in the
restructured representation (the solid line). The ordinate is the
absolute values of each coefficient $c_{i'}$ while the abscissa denotes
the basis number $i'$ reordered so that $|c_{i'}| \geq |c_{j'}|$ holds for 
$i' < j'$.}
\end{center}
\end{figure}

\begin{figure}[p]
\begin{center}
\scalebox{0.5}{\includegraphics{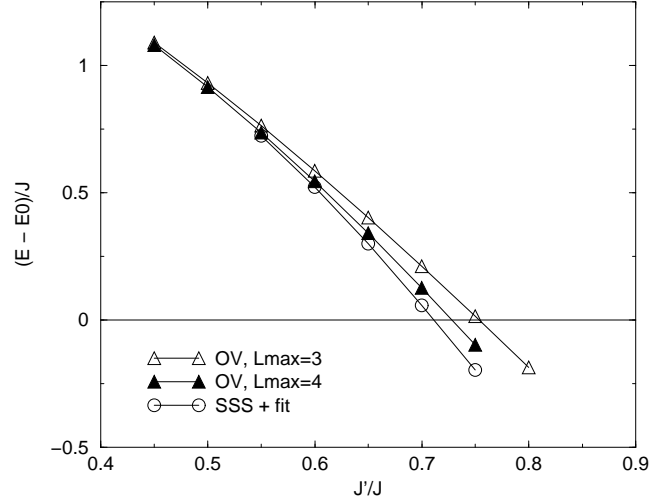}}
\caption{Results for $S = 0$ energy difference for the $8 \times 8$ SS model
obtained by the Stochastic State Selection (SSS) method with the 
 fit and the Operator Variational (OV) method\cite{ov}. Here $E_0=-24.0 J$ is
 the ground state energy in the singlet dimer phase.
}
\end{center}
\end{figure}
\begin{figure}[b]
\begin{center}
\scalebox{0.5}{\includegraphics{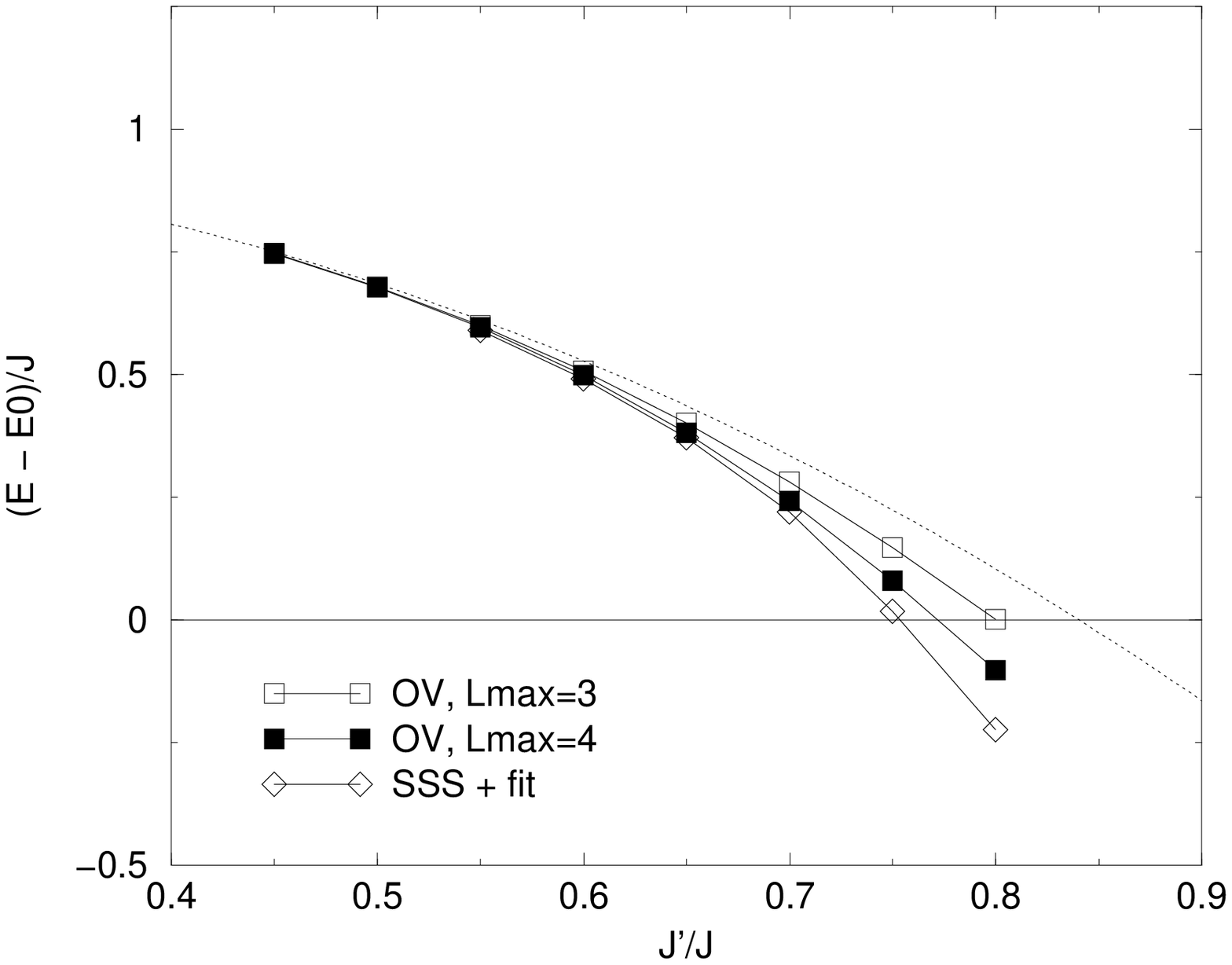}}
\caption{Results for $S = 1$ energy difference for the $8 \times 8$ SS model
obtained by the SSS method with the fit and the OV method\cite{ov}.
The dotted line is the spin gap estimated by the perturbation theory up
to the fifth order\cite{fuku}.}
\end{center}
\end{figure}


\begin{thebibliography}{99}
\bibitem{foot0}
For each $i$, all of $\eta_i^{(m)}$ $(m=1,2,\cdots,L+1)$ are generated
by the on-off probability function (\ref{pi}) with the {\em same} $a_i$
in (\ref{ai}), which is related to the expansion of the {\em initial}
state (\ref{expnd0}).
\bibitem{foot1}
For example, we observe that the $r$-ratio of the $8 \times 8$ SS system with 
$\beta J = 4$ in the Suzuki-Trotter method, where $\beta $ denotes the 
inverse temperature, is 
$r \sim 0.7$ at $J'/J=0.5$ while it drops to $r \sim 0.15$ at
$J'/J=0.635$ even in the restructured representation.
%
Here the $r$-ratio is defined by 
$r \equiv (N_+ - N_-)/(N_+ + N_-)$, $N_+$ ($N_-$) being 
the number of configurations with positive (negative) weight, 
respectively.
\end{thebibliography}
\end{document}